\documentclass[%
 reprint,
 groupedaddress,
 unsortedaddress,
 amsmath,amssymb,
 aps,
]{revtex4-2}

\usepackage{graphicx}% Include figure files
\usepackage{dcolumn}% Align table columns on decimal point
\usepackage{bm}% bold math
\usepackage{color, soul}

\begin{document}

%\preprint{APS/123-QED}

\title{Reference-enhanced X-ray Single Particle Imaging}

\author{Kartik Ayyer}
\email{kartik.ayyer@mpsd.mpg.de}
\affiliation{Max Planck Institute for the Structure and Dynamics of Matter, Luruper Chaussee 149, 22761 Hamburg, Germany}
\affiliation{Center for Free-Electron Laser Science, Luruper Chaussee 149, 22761 Hamburg, Germany}
\affiliation{The Hamburg Center for Ultrafast Imaging, Universit{\"a}t Hamburg, Luruper Chaussee 149, 22761 Hamburg, Germany}


\begin{abstract}
X-ray single particle imaging involves the measurement of a large number of noisy diffraction patterns of isolated objects in random orientations. The missing information about these patterns is then computationally recovered in order to obtain a three-dimensional structure of the particle. While the method has promised to deliver room temperature structures at near-atomic resolution, there have been significant experimental hurdles in collecting data of sufficient quality and quantity to achieve this goal. This paper describes two ways to modify the conventional methodology which significantly ease the experimental challenges, at the cost of additional computational complexity in the reconstruction procedure. Both these methods involve the use of holographic reference objects in close proximity to the sample of interest, whose structure can be described with only a few parameters. A reconstruction algorithm to recover the unknown degrees of freedom is also proposed and tested with toy-model simulations.
\end{abstract}

\maketitle

%\tableofcontents

\section{\label{sec:intro}Introduction}
Single particle imaging (SPI) at X-ray free electron lasers (XFELs) should, in principle, be able to image the structure and dynamics of biomolecules at near-atomic resolution and sub-ps timescales~\cite{Aquila:2015}. 
%Unfortunately, this promise has not yet been achieved due to various experimental difficulties. 
Challenges still remain in collecting sufficient number of high quality diffraction patterns, where high quality refers to diffraction patterns with low background and high signal, enough to enable the orientation determination and merging of individual patterns into a 3D structure. Various studies have been performed on the minimum quality of patterns that are still tolerable~\cite{Loh:2009,AyyerG:2015,Giewekemeyer:2019,Ayyer:2019}, and they conclude that single proteins can be imaged with currently available XFEL sources as long as the background scattering is significantly less than the scattered signal from the particle and that $10^5-10^6$ patterns from identical objects can be collected. Most experimental work~\cite{Ekeberg:2015,Rose:2018,Lundholm:2018} has been focussed on method development on much larger particles which scatter enough to be comfortably over the theoretical bounds.

Various techniques have been employed to deliver the samples into the X-ray focus. Aerosol methods have the lowest background~\cite{Munke:2016,Bielecki:2019}, but the particle densities are often so low as to make collection of a large number of patterns infeasible. One can collect more patterns by using a larger X-ray focus, but this proportionally reduces the scattered signal per pattern which means that the integrated signal with time stays constant. 

Alternatively, one can use a carrier medium for the particles, which can significantly increase the data collection rate. This medium can either be a liquid jet~\cite{Chapman:2011,Sierra:2012} or a solid substrate which is scanned in the X-ray focus~\cite{Hunter:2014,Nam:2016,Seuring:2018}. The scattering from the carrier medium unfortunately overpowers the signal from the particle, usually making even hit detection of single biomolecules impossible. This can, in principle, be improved by reducing the focus size significantly such that it almost matches the particle size. In that case, only a very small volume of the carrier medium will be illuminated, which should make the signal from the particle detectable. However, X-ray optics capable of such small foci and high flux densities at XFELs do not exist yet. 

In this paper, we discuss two alternative strategies to obtain high quality diffraction patterns with minimal modifications to currently available sample preparation and delivery technologies. The general principle for both of them is to gain signal-to-noise by including scattering from a strongly scattering reference~\cite{Boutet:2008,Lan:2014}. This is, of course, the holographic principle which has already been applied in diffractive imaging settings, notably in the form of Fourier transform holography~\cite{McNulty:1992} or as `free-flying' holography~\cite{Gorkhover:2018}. In both these cases, the stated goal has been to recover the structure of the particle in single shots without the need for phase retrieval. In contrast, the objective here is to recover the full 3D structure of a mostly reproducible object from a large number of patterns of composite structures consisting of the target object as well as a reference.  

The first composite object we consider is where a gold nanoparticle (preferably a sphere) is chemically attached to the target object in an aerosol imaging setup. The second system is to place a 2D crystal in the beam path with a unit cell comparable to the target object size. This can be achieved on a substrate in a straightforward manner by placing the 2D crystal on one side of the substrate and the sample on the other.

The common feature of these methods is that they add heterogeneity to the dataset, since the diffraction patterns vary not only in the orientation of the particles in the beam, but also in the properties and relative position of the reference. Composite objects like those we will discuss in Section~\ref{sec:singleth} have been proposed before~\cite{Shintake:2008} but this structural variability has been ignored and the reference and the target need to be separated by distances larger than the size of either, which is not the case here.

As we will see, in the methods proposed here, we gain experimental efficiency at the cost of computational complexity. In the next sections, we will discuss the two types of systems in detail. We will also describe a reconstruction algorithm to reconstruct the structure of the samples from these holographic patterns by treating these additional latent variables in a similar way as one does the unknown orientations in conventional SPI. For the nanoparticle reference case, we will also show the results of some 2D simulations on a toy model to show the efficacy of the algorithm.

In the following discussion, for convenience we refer to an identical or reproducible target object. One should note that exact, atomic resolution reproducibility is not required. The problem of conformational variability is the same faced by conventional SPI and the techniques being developed to deal with structural variability should also be applicable to the imaging methods described here.

\section{\label{sec:singleth}Single particle reference}
For the first holographic system, we consider a situation where the unknown target particle is attached to a single reference structure, specifically a spherical gold nanoparticle (AuNP). This reference has the benefits of alleviating problems with finding the hits over background due to the high scattering cross section, thus enabling the use of smaller particles than what could be used in conventional SPI. Secondly, due to the high density, the acceleration of the particles in the flow field is lower and the density of particles in the aerosol stream is higher, increasing the hit rate i.e. the fraction of pulses for which a particle is in the X-ray focus. Finally, these spherical references have just a single parameter to describe the structure, the radius. Gold nanospheres of a wide range of sizes are relatively easy to produce and are even commercially available. Various methods to link them to proteins and DNA have also been extensively studied~\cite{Aubin-Tam:2008,Mirkin:1996,Jones:2015}. However, these experimental benefits come at the cost of increased heterogeneity. 

In addition to the inherent structural variability of the target, we will have to solve for the relative positions of the reference and the unknown object and the size of the reference. If the reference was anisotropic and not a sphere, one would also have to contend with the relative orientation of the two objects, consequently making spheres even more desirable. Thus, we have 4 additional degrees of freedom for spherical references and more for an arbitrary one. However, we should note that not all of these degrees need be independent. Since the spheres are linked to points on the surface, there is a strong correlation between the position of the center and the size. Nevertheless, there is a substantial increase in the phase space of parameters that need to be solved for each diffraction pattern.

\begin{figure}
\centering
\begin{tabular}{ c c }
    \includegraphics[width=0.45\columnwidth]{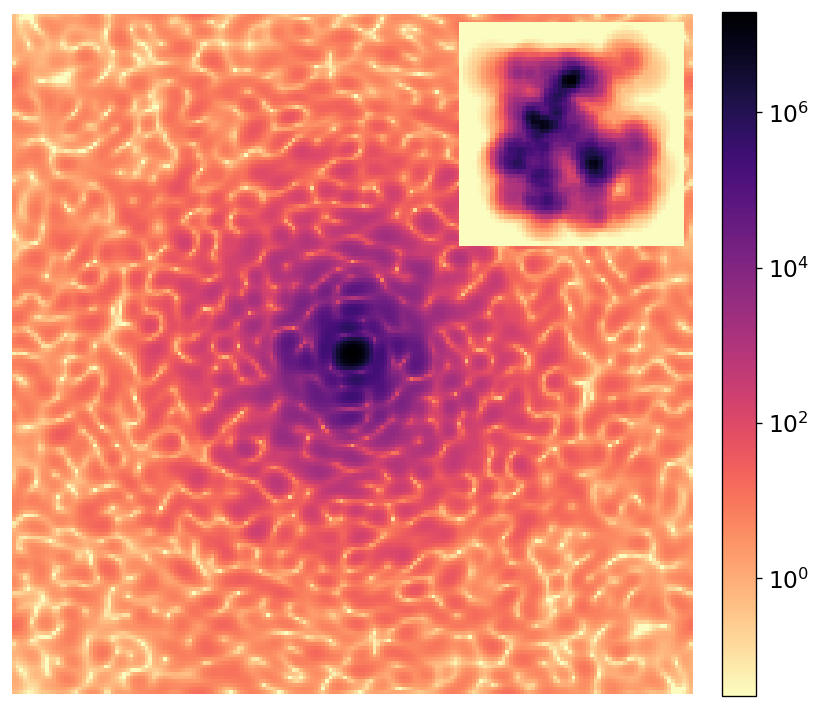} &  \includegraphics[width=0.45\columnwidth]{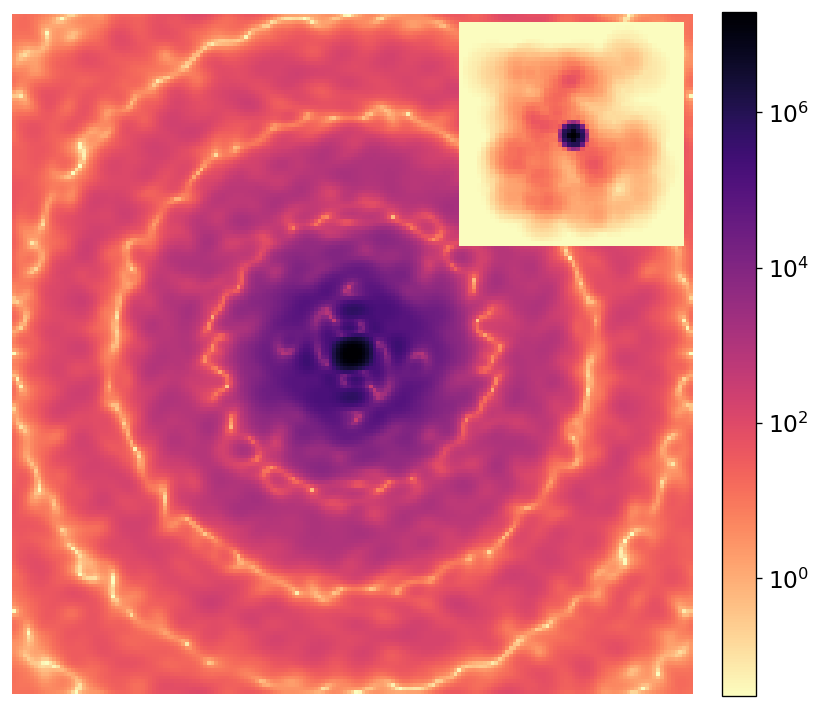} \\
    (a) & (b)
\end{tabular}
\caption{Random sphere cluster used as test object for illustration. (a) Intensity distribution of the test object shown on a logarithmic scale. Inset shows the projected electron density on a linear scale. (b) The same test object with a strongly scattering reference sphere attached. The main figure again shows the log-scale intensity distribution while the inset shows the projected electron density.}
\label{fig:singleobj}
\end{figure}

If one was performing a conventional SPI experiment with such samples, while the data collection process would be considerably eased by the experimental benefits described above, one would need to find a subset of patterns corresponding to the same composite object so that a 3D structure could be retrieved. This means throwing away a lot of data in order to find this subset. The holographic approach would be to decompose the composite object as the sum of the density of the spherical AuNP, $\rho_s(\mathbf{r}, d)$, and the unknown object $\rho_o(\mathbf{r})$, where $d$ represents the diameter of the sphere. The total electron density is
\begin{equation}
    \rho(\mathbf{r}) = \rho_o(\mathbf{r}) + \rho_s(\mathbf{r}-\mathbf{t}, d)
\end{equation}
where $\mathbf{t}$ is the relative shift of the centers of the two objects. The 3D intensity distribution of this object sampled in a single shot then becomes
\begin{equation}
    I(\mathbf{q}, d, \mathbf{t}) = \left|F_o(\mathbf{q}) + F_s(\mathbf{q}, d) e^{2\pi i \mathbf{q}.\mathbf{t}}\right|^2
    \label{eq:intens_holo}
\end{equation}
where the $F$ terms represent the Fourier transform of the densities and the shift of the sphere becomes a phase ramp in 3D. The Fourier transform of a sphere is straightforward to calculate analytically
\begin{equation}
    F_s(\mathbf{q}, d) \propto d^3 \left(\frac{\sin(s) - s\cos(s)}{s^3}\right)
\end{equation}
with $s=\pi d |\mathbf{q}|$. This is illustrated in Fig.~\ref{fig:singleobj} where one can see effect on the intensity distributions due to the addition of a spherical AuNP on a randomly generated organic-like cluster.

Equation~\ref{eq:intens_holo} makes it explicit that one must solve for the diameter and relative shift of each pattern in order to recover the structure of the common object. Unlike in the conventional SPI method, all diffraction patterns contribute to the structure, increasing the signal-to-noise ratio (SNR) and generating a higher resolution structure. Of course, the best case scenario would still be when the parameters $d$ and $\mathbf{t}$ have very narrow distributions, but this method effectively makes the experiment more tolerant to variations in the attachment process while still benefiting from the experimental benefits of the gold reference.

For a single shot, the noise at a given pixel is the square root of expected intensity from Eq.~\ref{eq:intens_holo} if we can reliably count the number of photons/pixel and there is not too much background. The signal is the difference compared to the sphere diffraction pattern. Thus, the SNR can be written as
\begin{equation*}
\mathrm{SNR}(\mathbf{q}) = \frac{I(\mathbf{q}, d, \mathbf{t}) - |F_s(\mathbf{q}, d)|^2}{\sqrt{I(\mathbf{q}, d, \mathbf{t})}}
\end{equation*}
For those pixels where the sphere signal $F_s(\mathbf{q})$ is much larger than that of the object, which is for most of the detector except near the sphere diffraction minima, the SNR simplifies to $2 |F_o(\mathbf{q})|$ which is a factor 2 higher than in the absence of a coherent reference ($F_s(\mathbf{q},d) = 0$). But more importantly, when the reference is strong, the SNR also becomes less sensitive to background scattering. 

\subsection{\label{sec:singlealg}Reconstruction algorithm}
The data set described above contains diffraction patterns which are noisy Ewald-sphere slices through many 3D intensities described by Eq.~\ref{eq:intens_holo} at a random, unknown orientation and scale factor, due to variations in the incident fluence. A reconstruction algorithm to recover the parameters of each pattern and the structure of the object is described in this section.

The EMC algorithm~\cite{Loh:2009} used widely in conventional SPI~\cite{Ekeberg:2015,Lundholm:2018,Rose:2018,Ayyer:2019} is composed of three steps in each iteration: expand, maximize and compress. The goal in each iteration is to find a model which has a higher likelihood of generating the data measured on the detector. The expand step is a transformation from model space to detector space for a given set of sampled hidden parameters. In the standard use case, the model is a grid of 3D intensities and the hidden parameters are the orientation. So in the expand step, one interpolates the 3D intensities along an Ewald sphere surface rotated by the given orientation and then applies standard polarization and solid angle corrections to produce the predicted intensities on the detector.

The maximize step finds an update to each of these detector views using the expectation maximization procedure and given a noise model. Usually, one also needs to find the maximum likelihood fluence factors. The result is a set of updated views which together have a higher likelihood, but are not necessarily consistent with a single 3D intensity. At the end of each iteration, this consistency is enforced in the compress step. The straightforward solution is to reinterpolate the detector views into the 3D model after undoing the detector corrections. Once the 3D intensity has converged, standard iterative phase retrieval algorithms are used to get the electron density. 

In the holographic case, the maximize step is left unchanged, since the objective is still to find the best possible predictions for the intensity at each detector pixel. The common 3D model is now not the 3D intensity of the whole object, but the complex Fourier transform of the unknown target, $F_o(\mathbf{q})$. In the expand step, one now interpolates the complex values along the Ewald sphere as before, but then converts them to detector intensities according to Eq.~\ref{eq:intens_holo} before applying detector corrections. As stated before, the predicted detector intensities depend upon the orientation, sphere diameter and relative shift.

The compress step, though, is not so straightforward, since the determination of the optimal $F_o(\mathbf{q})$ from many different detector intensities is effectively a phase retrieval problem. The first part of this is to recover the 3D intensities for a given set of $d$ and $\mathbf{t}$ diameter and shift parameters. This can be accomplished simply by interpolation as before. One is then left with many 3D intensity volumes, each corresponding to a different realization of Eq.~\ref{eq:intens_holo}, from which a single complex $F_o(\mathbf{q})$ must be determined. A \textit{divide and concur} difference map approach~\cite{Gravel:2008,Elser:2007} will be used here to solve this problem. 

Iterative projection algorithms like difference map~\cite{Elser:2003}, hybrid input-output (HIO)~\cite{Fienup:1978} etc. are used to solve constraint satisfaction problems like phase retrieval by searching for the intersection of two constraint sets in a high dimensional space. In these methods, update rules are composed of projections to sets, which are defined to be the point in the set closest to any given point in this space. The \textit{divide and concur} method extends these algorithms to an arbitrary number of constraint sets by expanding the state vector. If there are N constraints to satisfy, the new state vector is N copies of the original one. In the divide projection, each of the copies is projected to one of the constraint sets. The concur projection enforces consistency and the projection is just to replace each copy by the average over all of them. 

As applied to the compress step here, the divide projection will be a standard modulus projection from phase retrieval for each of the 3D intensity volumes. If the $n$-th intensity is $I_{\mathrm{obs},n}(\mathbf{q})$, the divide projection for that copy $F_{o,n}(\mathbf{q})$ will be
\begin{equation}
    \mathcal{P}_D[F_{o,n}(\mathbf{q})] = \sqrt{\frac{I_{\mathrm{calc},n}(\mathbf{q})}{I_{\mathrm{obs},n}(\mathbf{q})}} F_{\mathrm{calc},n}(\mathbf{q}) - F_s(\mathbf{q}, d_n) e^{2\pi i \mathbf{q}.\mathbf{t}_n}
\end{equation}
where $I_{\mathrm{calc},n}(\mathbf{q}) = |F_{\mathrm{calc},n}(\mathbf{q})|^2 = I(\mathbf{q}, d_n, \mathbf{t}_n)$ from Eq.~\ref{eq:intens_holo}. The concur projection will set each copy equal to the average over all of them. In addition, one can add additional real-space constraints like positivity or a bounded support and the projection will be to project the averaged copy to those constraints. This is especially helpful at low resolution where the phase shift due to the range of translations can be small. After convergence, the solution chosen for the next iteration is taken to be the concur projection, which is just the average over all copies with the real-space constraints applied.

Practically, it may often be the case that one can determine the sphere diameter from single shots to higher than the sampling precision in the EMC reconstruction. This is because the diameter can be estimated by the azimuthally averaged intensity $I(|\mathbf{q}|)$ which will have relatively good SNR even with only a few hundred scattered photons. In such a situation, the maximize step can be simplified to not calculate the probabilities over all diameters for every pattern, but just over the shift parameters.

\subsection{\label{sec:singlesim}2D simulations}
Simulations have been performed to illustrate the data produced and to demonstrate the reconstruction algorithm. For simplicity, a 2D toy model has been used which is rotated in-plane, similar to previous experiments to test the performance of the EMC algorithm with sparse data~\cite{Philipp:2012,Giewekemeyer:2019}. There is one parameter for the angle and two for the shift, but the qualitative structure of the problem remains the same. The test object representing the projected density of a random agglomeration of spheres and its Fourier intensity is shown in Fig.~\ref{fig:singleobj}(a).

\begin{figure}
\centering
\begin{tabular}{ c c }
     \includegraphics[width=0.45\columnwidth]{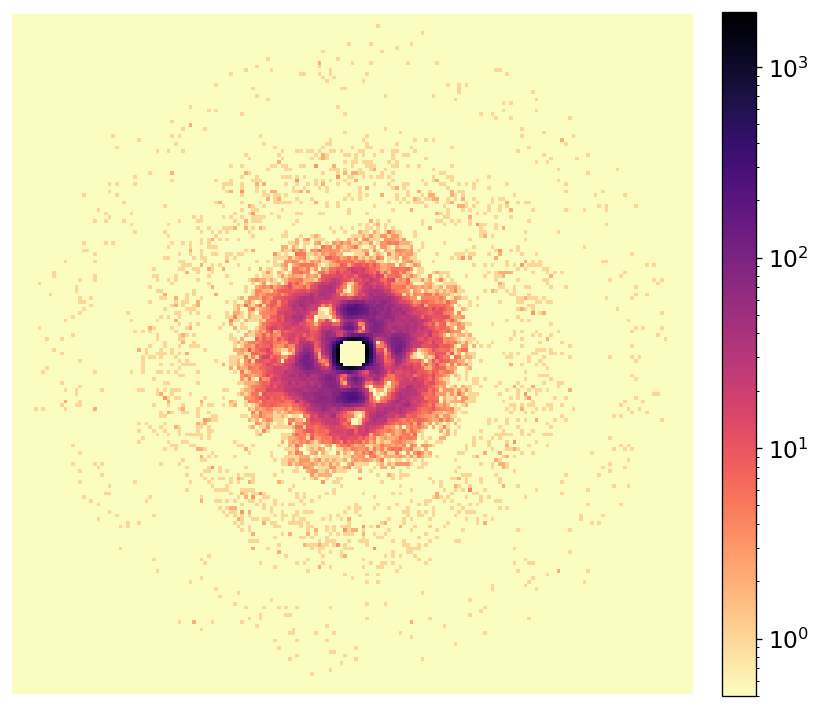} & \includegraphics[width=0.45\columnwidth]{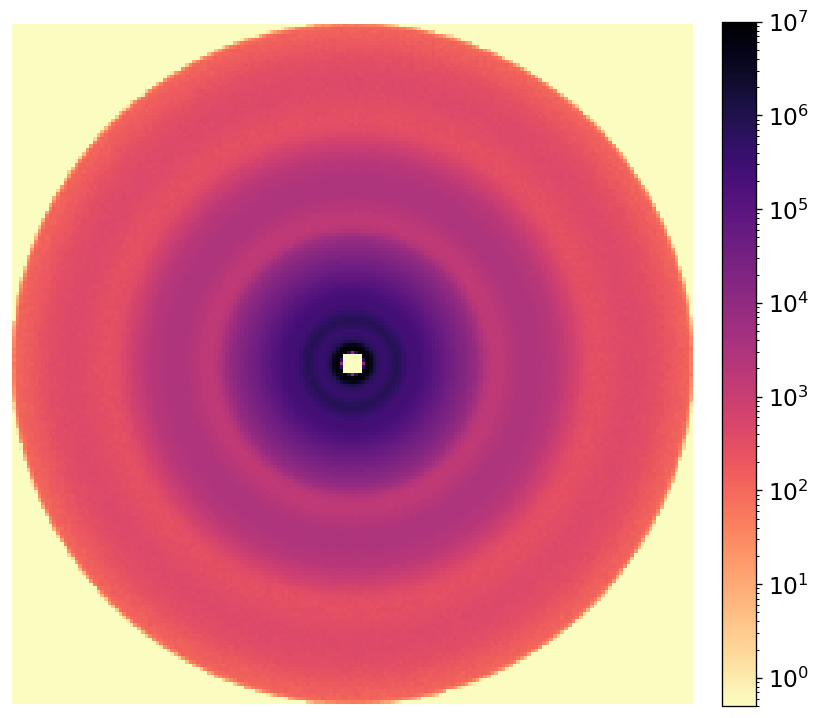} \\
     (a) & (b)
\end{tabular}
\caption{Illustration of forward calculation, used both to generate data as well as in the expand step. (a) Poisson sampled photon counts of the intensity distribution in Fig.~\ref{fig:singleobj}(b) shown on a logarithmic scale. Almost all the photons are concentrated at low resolution, as is expected from the Fourier transform of a compact object. The actual data will be a randomly rotated version of this pattern. (b) Virtual powder pattern, or integrated image for 10000 repeats of this process with different sphere diameters, positions and in-plane rotations. The innermost region and the corners of the detector were masked out.}
\label{fig:datagen}
\end{figure}

In order to generate the holographic data, the density of a sphere was added to that of the test object with the sphere center and diameters randomly sampled from normal distributions of a certain width. The result of one instance of this is shown in Fig.~\ref{fig:singleobj}(b), which also shows the intensity distribution of the composite object. These intensities were then Poisson sampled to generate photon counts per pixel (Fig.~\ref{fig:datagen}(a)) and then rotated in-plane by a random angle. For this simulation, 10000 patterns with $10^5$ photons/frame were generated. The sum of all the patterns, showing azimuthal symmetry due to random in-plane rotations, is shown in Fig.~\ref{fig:datagen}(b). The electron density of the sphere was chosen to be around 11 times that of the object, corresponding to the scattering factor ratio between gold and a protein-like material.

The sphere diameters for each shot were random sampled from a normal distribution with a mean of 7~pixels and a standard deviation of 1~pixel. For comparison, the test object image in the inset of Fig.~\ref{fig:singleobj}(a) is 50x50~pixels in size. The shift of the sphere center was randomly sampled from a 2D normal distribution with a standard deviation of 1~pixel. For these simulations, all of these parameters were independently generated, but as mentioned earlier, it is quite possible that the sphere diameter and center positions are correlated. The reconstruction algorithm could be made more efficient if these correlations were known.

The initial guess for the iterate, $F_o(\mathbf{q})$, is a set of random complex numbers. The reconstruction proceeds iteratively as described in Section~\ref{sec:singlealg} with the main difference that the object is 2D and there is only one degree of freedom for the orientations, namely the in-plane angle. Additionally, a support constraint is applied in conjunction with the \textit{concur} projection. The initial support is taken to be a 37x37 pixel square region centered in the field of view. The support is updated every 5 iterations using a \textit{Shrinkwrap}-like~\cite{Marchesini:2003} update rule where the current iterate is convolved with a Gaussian kernel of standard deviation 2 pixels and thresholded such that 2050 pixels are inside the support. 50 iterations of divide-and-concur difference map were applied for every EMC iteration with the $\beta$ parameter set to 1.

\begin{figure}
\centering
\begin{tabular}{ c }
    \includegraphics[width=0.9\columnwidth]{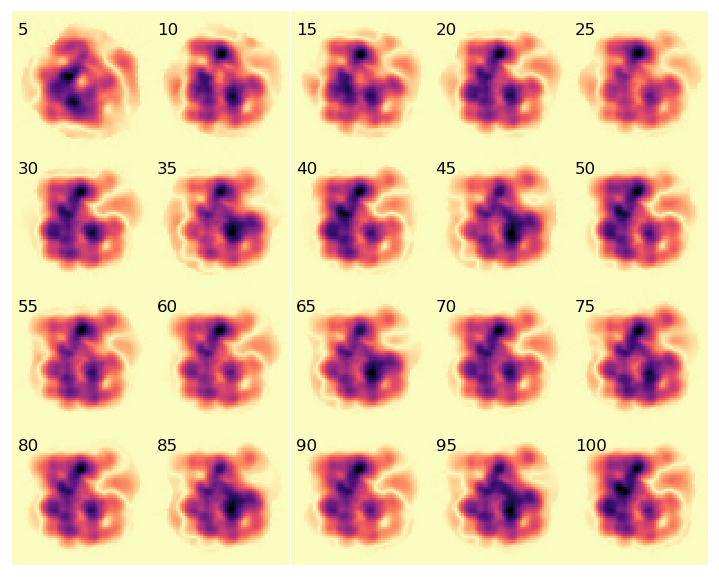} \\
    (a) \\
    \includegraphics[width=0.8\columnwidth]{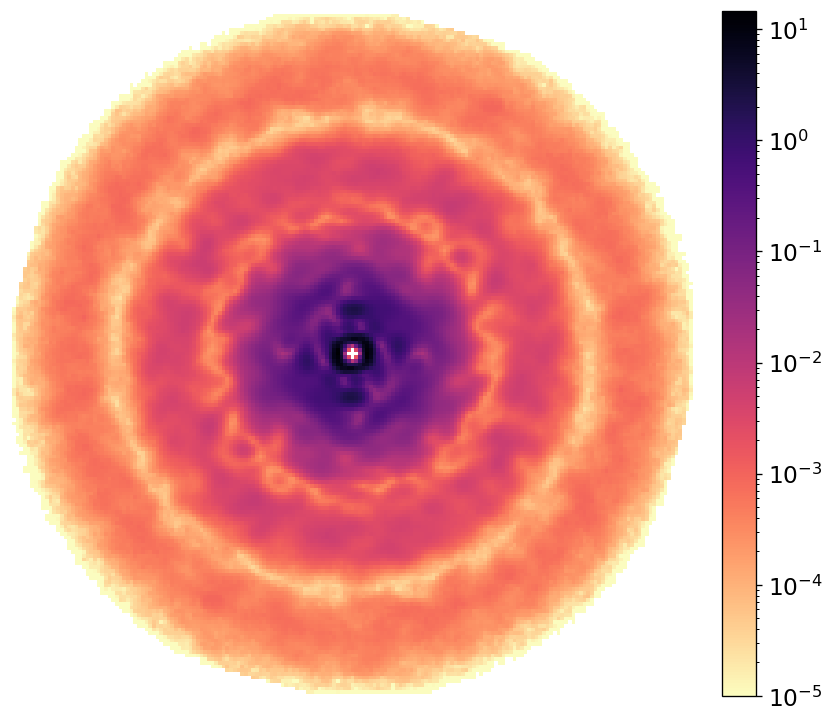} \\
    (b) \\
\end{tabular}
\caption{Single particle reference simulation results. (a) Reconstructed iterates after every 5 iterations. The reconstructions were rotated by $-15^\circ$ to facilitate comparison with the original image. (b) Intensity reconstruction of the final iteration with a sphere of diameter 7 nm and shifts of +0.5 pixels in both X- and Y-directions shown on a logarithmic scale.}
\label{fig:singleres}
\end{figure}

The results for a typical run are shown in Fig.~\ref{fig:singleres}. Figure~\ref{fig:singleres}(a) show the concur projection of the current iterate after every five iterations. These images were rotated by $-15^\circ$ to align with the true solution to make visual identification of features easier. The reconstruction will have, in general, a random rotational offset with respect to the ground truth. One can recognize that most of the structure of the test object has been recovered, but some additional density is also present. This can probably be optimized by modifying the phase retrieval parameters, especially those related to the support update. After every iteration, the 2D detector intensities were reconstructed for every set of sphere diameter and shift parameters by averaging over all the in-plane rotations. One of these is shown for the final iteration in Fig.~\ref{fig:singleres}(b). This can be compared with the true intensities with similar parameters shown in Fig.~\ref{fig:singleobj}(b).

\begin{figure}
\centering
\begin{tabular}{ c }
    \includegraphics[height=0.7\columnwidth]{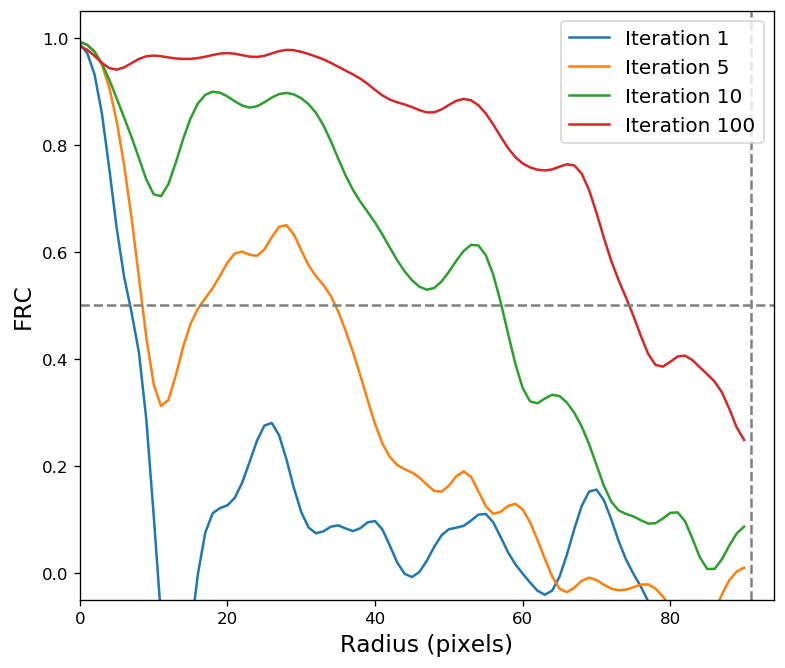} \\
    (a) \\
    \includegraphics[height=0.75\columnwidth]{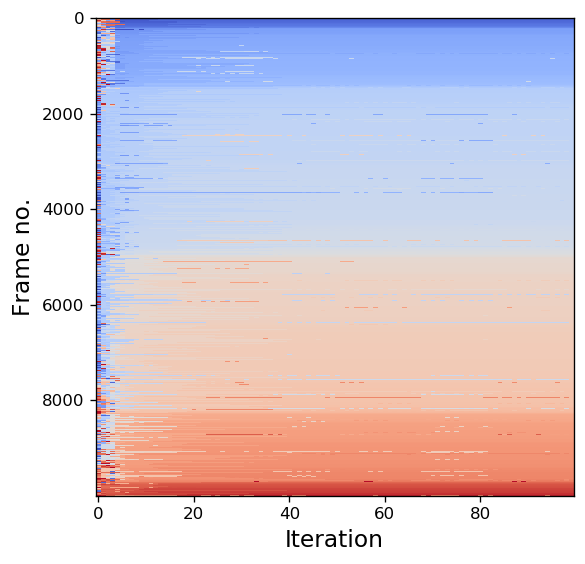} \\
    (b) 
\end{tabular}
\caption{Single particle reference simulation metrics. (a) Fourier ring correlation between reconstructions and ground truth as a function of iteration number. (f) Convergence plot of most likely parameters for each frame as a function of iteration.}
\label{fig:singlemetrics}
\end{figure}

Fig.~\ref{fig:singlemetrics}(a) shows the Fourier ring correlation (FRC) metric~\cite{Saxton:1982} comparing the reconstructions for a few iterations to the ground truth. The vertical dashed line indicates the edge of the `detector' and the horizontal dashed line indicates the somewhat arbitrary FRC$=0.5$ cutoff. The final plot Fig.~\ref{fig:singlemetrics}(b) shows the convergence of the most likely parameter (diameter, position, orientation) for each pattern as the iterations proceed. This convergence plot is the same one used in the \textit{Dragonfly}~\cite{Ayyer:2016} software and shows how after around 10 iterations already the most likely parameters are mostly converged.

\section{\label{sec:latticeth}Reference lattice}

\begin{figure}
\centering
\begin{tabular}{ c }
    \includegraphics[width=0.8\columnwidth]{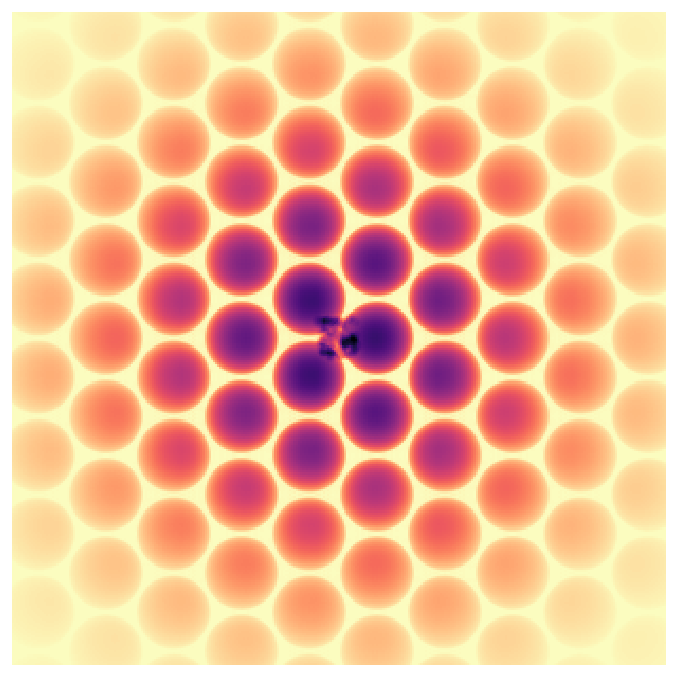} \\
    (a) \\
    \includegraphics[width=0.8\columnwidth]{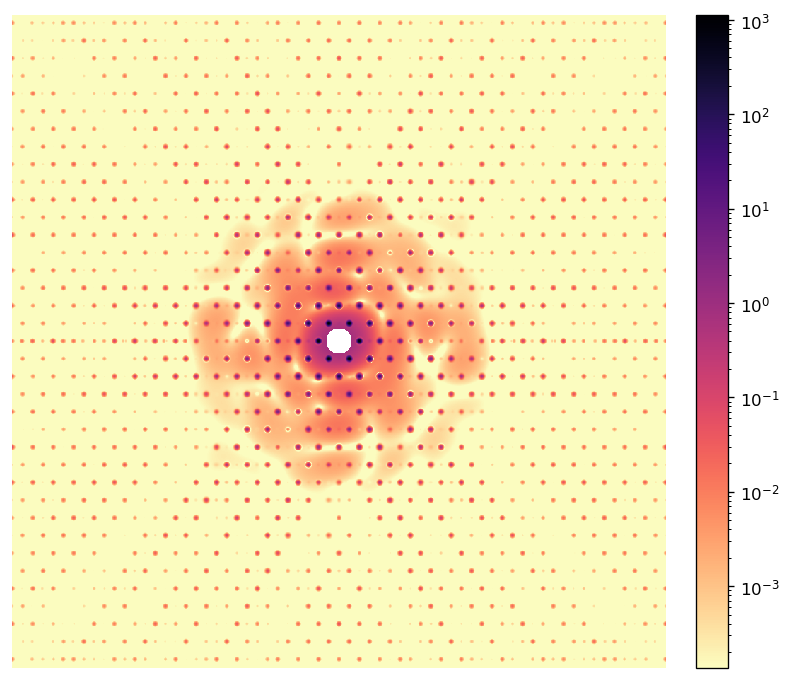} \\
    (b) \\
\end{tabular}
\caption{2D schematic showing the diffraction from a 2D crystal made up of spheres in a triangular lattice with the same cluster test object used in Section~\ref{sec:singlesim}. (a) The projected electron density showing the lattice, the target as well as the probe which here had a full-width at half-maximum of 5 unit cells. (b) The expected intensity distribution due to such a composite object on a logarithmic scale. The peak intensities are modulated by the orientation and position of the target object. One can also see the weak diffuse scattering from the molecular transform of the target object itself, but this will likely be drowned in background scattering from the substrate.}
\label{fig:latticeexp}
\end{figure}

The second method we will discuss to provide a holographic reference is to utilize a 2D crystal, either patterned onto a chip or as a self-assembled colloidal crystal~\cite{Vogel:2015}. An illustration of the experimental data for this is shown in Fig.~\ref{fig:latticeexp}. One way to get such data is to have the 2D crystal on one side of a substrate and the target samples randomly dispersed on the other side. Such fixed target scanning geometries have been used for single particle imaging of gold clusters~\cite{Nam:2016} as well as 2D crystallography~\cite{Hunter:2014} and fiber diffraction~\cite{Seuring:2018}. As before, one would have to solve for additional parameters on top of the object orientation, namely the position of the object's center modulo the unit cell.

However, the big advantage of using a lattice reference compared to directly putting the sample on the substrate is the extreme gain in background-tolerance obtained by using integrated Bragg peak intensities. Since experimental background scattering from the substrate and other beamline components is slowly varying, it is often straightforward to determine the integrated peak intensities as is standard in crystallography. In contrast to the single particle reference discussed in Section~\ref{sec:singleth}, the 2D crystal is prepared separately from the target sample and the relative position and orientation of the two systems should be uniformly distributed.

Let the electron density of the unit cell be $\rho_c(\mathbf{r})$ and of the unknown object be $\rho_o(\mathbf{r})$ as before. Let the unit cell be larger than the object and the illuminated region represented by a probe function $P(\mathbf{r})$ be which is significantly larger than both. The first condition can be relaxed somewhat but is convenient for sufficient sampling, especially at low resolution, as will soon be evident. The second condition is necessary to avoid going into the regime of ptychography, where one would have to recover the shot-by-shot probe profile~\cite{Liu:2018,Sala:2017}.

The 2D crystal can be represented as the unit cell convolved with a grid of Dirac delta functions,
\begin{equation}
\rho_L(\mathbf{r}) = \rho_c(\mathbf{r}) * \sum_i \delta(\mathbf{r} - \mathbf{r}_i)
\end{equation}
where the $*$ symbol represents convolution. The scattering contrast is the sum of the electron densities of the crystal and the rotated and translated object multiplied by the probe
\begin{equation}
\rho(\mathbf{r}) = \left[\rho_L(\mathbf{r}) + \rho_o(\mathbb{R}.\mathbf{r} - \mathbf{t})\right] \cdot P(\mathbf{r})
\end{equation}
where $\mathbb{R}$ and $\mathbf{t}$ are rotations and translations of the object with respect to a canonical configuration.

The far-field diffraction pattern is the Fourier transform of $\rho(\mathbf{r})$ sampled along the Ewald sphere. Using the convolution theorem, we get 
\begin{equation}
F(\mathbf{q}) = \sum_i F_c(\mathbf{q}_i) F_P(\mathbf{q} - \mathbf{q}_i) + F_o(\mathbb{R}.\mathbf{q})e^{2\pi i \mathbf{q}.\mathbf{t}}
\end{equation}
where the $F_\_(\mathbf{q})$ terms represent the Fourier transforms of the corresponding real-space quantities. Since the spot size is assumed to be large compared to the object, the effect of convolving $F_o(\mathbf{q})$ is neglected. The first term represents a reciprocal lattice of broad Bragg peaks whose shape is given by the probe Fourier transform and the height by the magnitude of the unit cell transform at the center of the Bragg peak. The diffracted intensities are given by
\begin{align}
I(\mathbf{q}) &= F(\mathbf{q}) F^*(\mathbf{q}) \nonumber \\
\begin{split}
&= \sum_i \left|F_c(\mathbf{q}_i) F_p(\mathbf{q} - \mathbf{q}_i)\right|^2 + \left|F_o(\mathbb{R}.\mathbf{q})\right|^2 \\
&\quad + F_o(\mathbb{R}.\mathbf{q})e^{2\pi i \mathbf{q}.\mathbf{t}}\sum_i F_c^*(\mathbf{q}_i) F_p^*(\mathbf{q} - \mathbf{q}_i) + \mathrm{c.c.}
\end{split}
\end{align}
where $\mathrm{c.c.}$ refers to the complex conjugate of the previous term. The first term is simplified by the assumption that the probe is much larger than the unit cell, and thus the width of the Bragg peaks are much less than the reciprocal lattice constant. In practice, there will be background scattering from various components in the beamline added to the intensities. The background is measurably higher than the aerosol-based sample delivery method discussed in Section~\ref{sec:singleth}. As in serial crystallography, this can be mitigated by working with the integrated intensity of each Bragg peak at $\mathbf{q}_i$. The relatively slowly varying $|F_o|^2$ term is assumed to be lost in the background. Also, if the probe is much larger than a unit cell as assumed, we would expect the Bragg peaks to be much brighter than the diffuse molecular transform of the target object. If the integration of the probe function $F_p(\mathbf{q}-\mathbf{q}_i)$ in the neighbourhood of the peak is $N$, the integrated peak intensities are given by
\begin{align}
\begin{split}
I_\mathrm{obs}(\mathbf{q}) &= \left|N F_c(\mathbf{q}_i)\right|^2 \\
& \quad + 2 N \left|F_o(\mathbb{R}.\mathbf{q}_i)\right|\left|F_c(\mathbf{q}_i)\right|\cos(\phi_o + 2\pi \mathbf{q}_i.\mathbf{t} - \phi_c)
\end{split}
\label{eq:lattice_intens}
\end{align}
where $\phi_\_$ represent the phases of the Fourier transform terms. With the choice of a simple object for the unit cell, $F_c(\mathbf{q})$ can be pre-calculated or measured beforehand.

The reconstruction approach very similar to that in Section~\ref{sec:singlealg} is applicable, except that Eq.~\ref{eq:intens_holo} is replaced by Eq.~\ref{eq:lattice_intens} and the intensities are only sampled at the reciprocal lattice points. Depending on the relative sizes of the object and the unit cell, a worry might be that the sampling rate of the Bragg peaks may be insufficient to determine the structure \textit{ab initio}. However, with random orientations, the sampling provided by $\mathbb{R}.\mathbf{q}_i$ will be sufficient beyond the first few \textit{hk} orders. Nevertheless, for completeness at low resolution, a unit cell larger than the object would be preferable. 

The other experimental parameter that requires some consideration is the size of the beam focus $P(\mathbf{r})$ compared to the lattice constant. The biggest challenge in determining $F_o(\mathbf{q})$ is the determination of the translation and orientational parameters for each diffraction pattern. For variable translations, Eq.~\ref{eq:lattice_intens} can be seen as a constant plus a scaled cosine as a function of $(\mathbf{q}.\mathbf{t})$. The amplitude of the cosine term is the signal relevant to determining the translation, $\mathbf{t}$. The noise in the Poissonian photon counting regime is the square root of $I_\mathrm{obs}(\mathbf{q})$, which is approximately just the square root of the first term, $N|F_c(\mathbf{q}_i)$. Thus, the signal-to-noise ratio (SNR) is 
\begin{equation*}
\frac{2 N |F_o(\mathbb{R}.\mathbf{q}_i)||F_c(\mathbf{q}_i)|}{N|F_c(\mathbf{q}_i)|} = 2 |F_o(\mathbb{R}.\mathbf{q}_i)|
\end{equation*}
which is independent of $N$. Since background subtraction during peak integration is an additional source of noise, $N$ should be as large as possible. However, detectors lose the ability to count individual photons if the signal is too high, either due to saturation or due to switching to a lower gain mode. The noise in the measurement would then be higher than $\sqrt{I_\mathrm{obs}}$ because of the additional uncertainty introduced by not knowing how many photons were measured. Thus, the optimal probe size in the absence of background would be the largest $N$ where the detector can still count photons. This optimum would shift to larger $N$ when there is significant background, would likely be the limiting experimental factor, especially at high resolution.

%%%%%%%%%%%%%%%%%%%%%%%%%%%%%%%%%%%%%%%%%%%%%%%%%%%%%%%%%%%%%%%%%%%
% [ The paragraphs below are wrong. Kept as a curiosity ]
%%%%%%%%%%%%%%%%%%%%%%%%%%%%%%%%%%%%%%%%%%%%%%%%%%%%%%%%%%%%%%%%%%%
%Since N is relatively large, the noise is approximately equal to $N|F_c(\mathbf{q}_i)|$ in the photon counting regime. Thus, the signal-to-noise ratio (SNR) is proportional to $1/N^2$, suggesting that the probe must be as small as possible. From an SNR perspective, the optimal solution is to match the beam size to the object size but this may not be feasible, nor desirable due to the experimental and analytical complications, especially if there are shot-by-shot instabilities in the beam profile as is common with free electron lasers. 

%Additionally, strong Bragg peaks enable us to align patterns relatively easily and obtain signal at high resolution beyond where Bragg peaks are easily visible in single shots. An intriguing possibility for improving experimental efficiency is to purposefully have a slightly disordered crystal. The crystal transform $N F_c(\mathbf{q})$ will have a relatively high Debye-Waller factor compared to the target object and at a certain resolution, this will lead to a higher SNR. At an even higher resolution, the dominant term will become the object's molecular transform $F_o(\mathbf{q})$, which may possibly be averaged and extracted using the orientation determined from lower resolution peaks even if there is significant background.
%%%%%%%%%%%%%%%%%%%%%%%%%%%%%%%%%%%%%%%%%%%%%%%%%%%%%%%%%%%%%%%%%%%

\section{\label{sec:discussion}Discussion}
X-ray single particle imaging remains an experimentally demanding method to determine the structure of uncrystallized single biomolecules. Problems still remain in obtaining sufficient data of high quality and questions remain over feasibility in transitioning to smaller particles.

Two new methodologies have been proposed here, both of which improve experimental efficiency by incorporating strongly scattering holographic references, but add complexity because the composite object is not necessarily reproducible. A reconstruction algorithm involving a modification to the EMC algorithm is proposed for recovering the additional degrees of freedom. The key insight is to separate the reference and the object, as shown in Eqs.~\ref{eq:intens_holo} and \ref{eq:lattice_intens} and explicitly sample the different degrees of freedom introduced by the addition of the reference. These methods also differ from other commonly used holographic methods like Fourier transform holography or in-flight holography where the references are separated to such an extent that one can perform single-shot imaging without the need for phase retrieval.

The first reference proposed is where one chemically attaches a strong reference scatterer like a gold nanosphere to the target object in an aerosol imaging setup. The size and relative position of the sphere is assumed to vary shot-to-shot in some interval. The reference makes hit detection easier and improves the hit rate since the composite objects are denser, and hence slower in the aerosol stream. 2D simulations were performed showing the reconstruction process and the ability to determine the unknown degrees of freedom (sphere size, position and object orientation).

The second geometry uses of a 2D crystal reference in a scanning fixed-target sample geometry. High hit rates can be achieved by controlling the density of particles deposited on the surface. The lattice reference produces Bragg peaks in the diffraction pattern which are much more robust to background, which is usually a limiting issue due to the presence of a substrate in the beam path. The integrated peak intensity contains information about the structure of the target object as well as its position relative to the lattice unit cell. The gain in background tolerance may also enable sample preparation methods which are either easier or leave the biomolecule in a closer-to-native state, like liquid cells or graphene sandwiches.

Further work is required to test the limits of the method in terms of minimum target object size with currently available XFEL parameters. The author also hopes that these ideas will be tested experimentally in the near future, potentially opening up a new dimension in optimizing experiments to achieve the goal of atomic-resolution structure and dynamics of uncrystallized biomolecules.

\begin{acknowledgements}
The author wishes to acknowledge the extremely valuable discussions with Henry Chapman during the preparation of this manuscript.
\end{acknowledgements}

\bibliography{main}

\end{document}